\documentclass[12pt]{iopart}
\usepackage[T1]{fontenc}
\usepackage[utf8]{inputenc}

\usepackage{lmodern}

\usepackage[eulergreek]{sansmath}
\sansmath

\usepackage[onehalfspacing]{setspace}
\usepackage{microtype}
\DisableLigatures{}

\usepackage{graphicx}
\usepackage{cite}

\begin{document}

\title[Interaction of intense femtosecond laser pulses with argon
  microdroplets]{The interaction of intense femtosecond laser pulses with
  argon microdroplets studied near the soft x-ray emission threshold}

\author{R. Irsig$^1$, M. Shihab$^{1,2}$, L. Kazak$^1$ , T. Bornath$^1$,\\
  J. Tiggesbäumker$^{1,*}$ R. Redmer$^1$, and K.-H. Meiwes-Broer$^1$}

\address{$^1$ Institut für Physik, Universität Rostock, 18059 Rostock, Germany}
\address{$^2$ Department of Physics, Faculty of Science, Tanta University, Tanta 31527, Egypt}
\address{$^*$ E-mail :  josef.tiggesbaeumker@uni-rostock.de}

\begin{abstract}
The extreme ultraviolet plasma emission from liquid microsized argon droplets
exposed to intense near-infrared laser pulses has been
investigated. Emission from the warm dense
matter targets is recorded in a spectral range in between 16 and 30\,nm at laser intensities of $10^{14}$\,W/\,cm$^2$. Above the emission threshold, soft x-ray radiation exponentially increases with the pulse energy whereby a
strong dependence of the yields on the pulse duration is observed,
which points at an effective electron collisional heating of the
microplasma by inverse bremsstrahlung. Accompanying 
hydrodynamic simulations reveal the temporal and spatial development
of the microplasma conditions. The good agreement in between the measured and calculated emission spectra as well as the extracted electron temperatures confirm that hydrodynamic simulations can be applied in the analysis of strongly excited droplets.
\end{abstract}

\section{Introduction}
\label{intro}

The interaction of intense near-infrared (NIR) laser light with small particles
has been of particular interest in the last decades, since it allows to
study matter of finite size under extreme
conditions \cite{KrainovPR2002,SaalmannJPB2006,FennelRMP2010,McNaughtAPL2001,WielandAPB2001,ParraAPA2003,GarnovAIP2009}.
Due to the strong absorption capabilities, nanoparticles are
effectively transformed into nanoplasmas
\cite{DitmirePRL1997}, which results in the emission of highly charged
and energetic ions \cite{SnyderPRL1996,LeziusPRL1998,KoellerPRL1999}
and fast electrons \cite{SpringatePRA2003,PassigNJP2012}. The
generation of short wavelength radiation
\cite{McPhersonNat1994,IssacPP2004} and high harmonics
\cite{TischJPB1997,BulanovPP1994} allows to resolve further details of
the size-limited plasma conditions
\cite{McNaughtAPL2001,DitmirePRA1996,ZweibackPRA1999,ParraPRE2000,PrigentPRA2008}.

 With microdroplets, the dimension of the target becomes comparable to
NIR laser wavelength and propagation phenomena show up. The laser
light wave penetrates into the droplet up to a region where the
critical density in the plasma is achieved. In addition, the spatial
density scale length is much lower compared to, e.g., the Ti:sapphire
laser wavelength of 800\,nm. Hence, absorption of radiation occurs
only near the surface of the droplet. The resulting hot plasma
is far from equilibrium. As a 
consequence, the dynamics of the laser-produced plasma layer has to be
simulated in order to resolve details of the processes involved. The
microplasma may retroact on the light propagation, further influencing
the absorption properties including extreme ultraviolet (EUV) re-absorption. As a result of
strong electron density gradients, wave propagation phenomena show up,
e.g., shock waves \cite{Zeldovich1966}, wave breaking and
turbulence \cite{VarPRL12}. The limited size of the target space
charge confinement of hot electrons will lead to an efficient coupling
of the laser radiation into the
particle \cite{SperlingNJP2013}. Particle-in-cell simulations show that an
inhomogeneous plasma layer develops which extends deeper into the
droplets than expected from classical predictions~\cite{LisPRL13}.

Beside particle plasma diagnostics, radiation has proven to give
meaningful information to characterize the complex dynamics in
microplasmas \cite{GleRMP09,OstAiP13}.  For example, intense
short-wavelength radiation from extended-ultraviolet 
free-electron lasers can isochorically heat
particles, whereby self-Thomson scattering can be exploited
to characterize the resulting warm dense matter (WDM)
state \cite{FaePRL10}. EUV-pump--EUV-probe studies were successful in
mapping the solid-to-plasma transition in real
time \cite{ZasPRL14}. Combining time-resolved soft x-ray diagnostics
with microplasma generation by intense NIR laser pulses remains a
challenge due to the initially incomplete spatial heating. WDM targets also
represent laboratory systems to simulate conditions found in planets
and are thus relevant for astrophysical
research \cite{GuiS99}. Finally, the short wavelength emission from
particles in a molecular beam have potential for
applications like high repetition rate table-top x-ray
sources \cite{StielSPIE2002}, 
because the technique provides a regenerative target and nearly debris-free conditions.

In this contribution we address in a joint experimental and
theoretical project the generation of the transient plasma state by
analyzing the resulting EUV radiation from an NIR laser-induced argon
microplasma. In a simplified view of the interaction, ionization of the liquid microdroplets is triggered by tunnel ionization. As the electron quiver amplitude at
$10^{14}$\,W/\,cm$^2$ 
\cite{KrainovPU2007} exceeds the interatomic distances in the liquid
\cite{HenshawPR1953}, efficient collisional ionization into
high charge states will take place producing a plasma at near solid-state
density.  
The thin plasma sheet provides in the following for an efficient heating by
inverse bremsstrahlung leading to a strong nonlinear dynamics. After
the impact of the laser pulse, the outward expansion of the plasma
layer leads to a decrease in the plasma density and an adiabatic
cooling \cite{ShihabAPB2016}. Soft x-ray emission as a characteristic fluorescence
signature of the plasma conditions occurs on a pico- to nanosecond time scale.

In contrast to other experiments on droplets at near relativistic intensities, e.g. \cite{WielandAPB2001,ParraPRE2000}, 
moderate laser intensities are applied in order to study the dynamics close to the argon tunnel ionization threshold of $2\cdot10^{14}$\,W/\,cm$^2$ \cite{AugstPRL1989}. 
To solidify the investigation the experiments are accompanied by theoretical work, i.e the experimental average electron temperatures extracted from the
EUV spectra will be compared to simulations. The good agreement of the theoretical results to the experiments point out that the microplasma dynamics can be described well by 1D hydrodynamical codes. 

\section{Experiment}

\begin{figure}[t]
\centering
\includegraphics[width=0.5\textwidth]{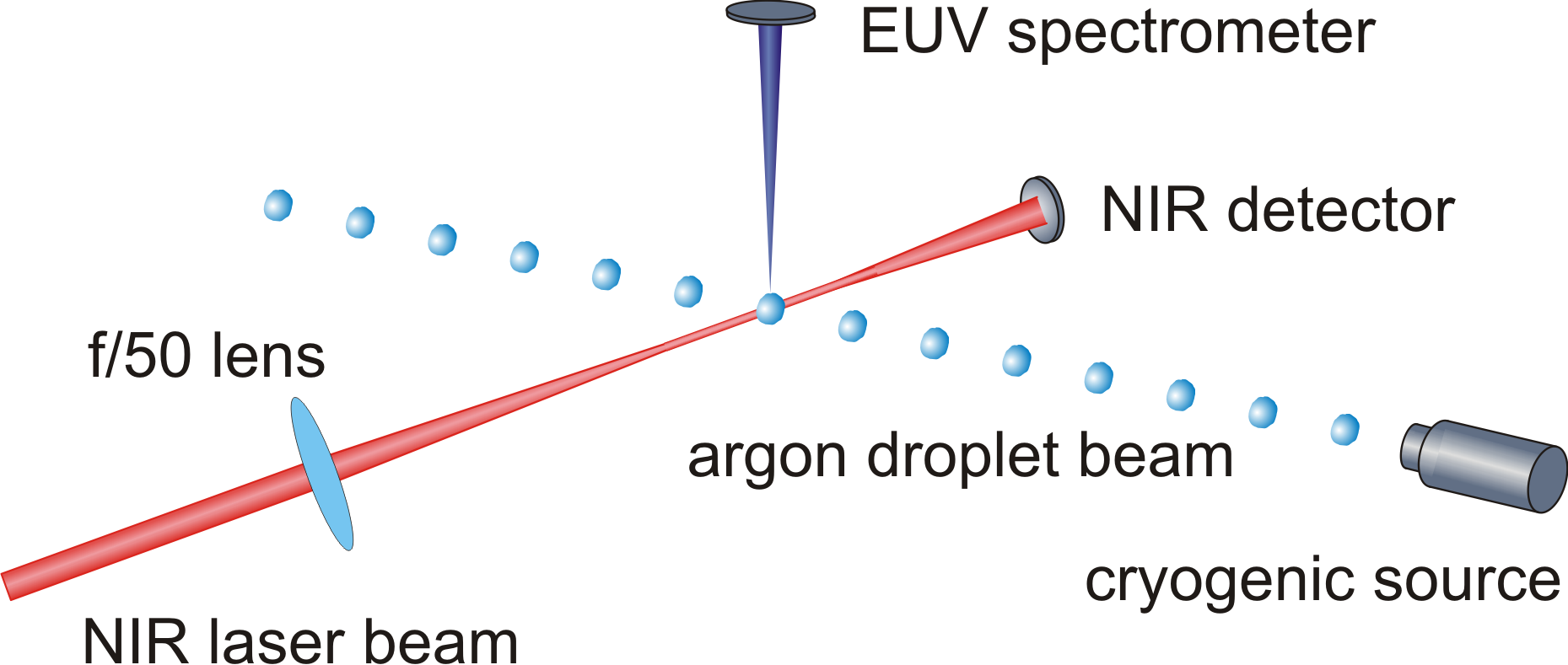}
\caption{(color online) Schematic view of the experimental setup. NIR femtosecond laser pulses at intensities of $10^{14}$\,W/cm$^2$ are focused onto a beam of argon
  microdroplets generated by a cryogenic source. In the strong-field
  interaction, a warm dense matter target is produced. Highly charged
  ions present in the microplasma emit characteristic short wavelength
  radiation. The EUV spectra are recorded between 16 and 30 nm (41
  to 77 eV) by a home-build flat-field spectrometer aligned
  perpendicular to the laser-droplet plane.}
\label{fig:exp_setup}       
\end{figure}

To effectively conduct the experiment, a regenerative high-density
target is utilized, i.e., a liquid jet source similar to the one used
in \cite{ToleikisJPB2010}. In short, the microdroplet beam is generated by
expanding liquefied argon gas (100\,K at 5\,bar) through a 5\,$\mu$m
nozzle into vacuum forming a continuous filament.  Due to surface
vibrations, the filament undergoes a Rayleigh breakup and splits into
monodisperse droplets \cite{ToenniesMP2013} with diameters of
approximately 9\,$\mu$m.  The droplets are irradiated by pulses from a
Ti:sapphire laser system with a center wavelength at 810\,nm and a pulse repetition rate of 1\,kHz.  The
laser beam is focused with a f/50 lens, see
Fig. \ref{fig:exp_setup}. Experiments are performed with pulse
durations between 100\,fs and 800\,fs and pulse energies up to
1.5\,mJ, which corresponds to a maximal pulse intensity of
$9\cdot10^{14}$\,W/cm$^2$ in the interaction region.  The EUV emission
produced in the strong-field interaction with the droplets is recorded
by a home-build EUV spectrometer aligned $90^{\circ}$ to both laser
and droplet beam. The spectrometer consists of an aberration corrected
flat-field grating as single optical element \cite{KitaAO1983}. A
back-illuminated CCD camera (Andor, model \textit{Newton 940 DO})
serves for EUV photon detection. A 200\,nm aluminum foil is used
to suppress scattered laser light. With the instrument, spectra in a range 
between 16 and 30\,nm can be recorded.   

As the Rayleigh breakup takes place spontaneously, 
a temporal synchronization between droplets and laser pulses cannot be achieved. 
In order to estimate the efficiency at which droplets are hit by laser radiation,
the transmitted energy of single pulses after passing the interaction
region is determined (NIR detector in Fig. \ref{fig:exp_setup}). From the analysis we found that about 70\% of the pulses interact with a droplet. To suppress signal fluctuations 
spectra are integrated over $10^4$ laser pulses.

\section{Simulation}

The interaction of intense femtosecond laser pulses with microdroplets is
studied using the radiation-hydrodynamic HELIOS code
\cite{MacFarlaneJQSRT2006}.  Briefly, HELIOS provides a Lagrangian
reference frame where electrons and ions are assumed to be
co-moving. Pressure contributions to the equation of motion stem from
electrons, ions and radiation.  Separate ion and electron
temperatures and flux-limited Spitzer thermal conductivity are
assumed.  Deviations from local thermodynamic equilibrium conditions are
accounted for by solving multi-level atomic rate equations at each
time step in the simulation.  The laser energy is deposited via
inverse bremsstrahlung as well as bound-bound and bound-free
transitions using an SESAME-like equation of state.  The EUV emission
of the microdroplets is calculated using SPECT3D based on the
temperature and density of the plasma plume expected via
radiation-hydrodynamic simulations.  SPECT3D is a
collisional-radiative code \cite{MacFarlaneHEDP2007} whereas the
radiation incident at a detector is estimated by solving the radiative
transfer equation along a series of lines-of-sight through the plasma
grid. Atomic cross section data are generated using a collection of
publicly available codes
\cite{WangPhD1991,FischerCPC1978,Abdallah1988,WangRSI1992,WangPRE1993}, 
where different processes are considered, such as excitation, electron-impact ionization, 
photoionization, radiative recombination, autoionization, and dielectronic recombination. In addition,
continuum lowering effects are implemented using an occupation probability
model.

\section{Experimental results}

\begin{figure}[t]
\centering
\includegraphics[width=0.5\textwidth]{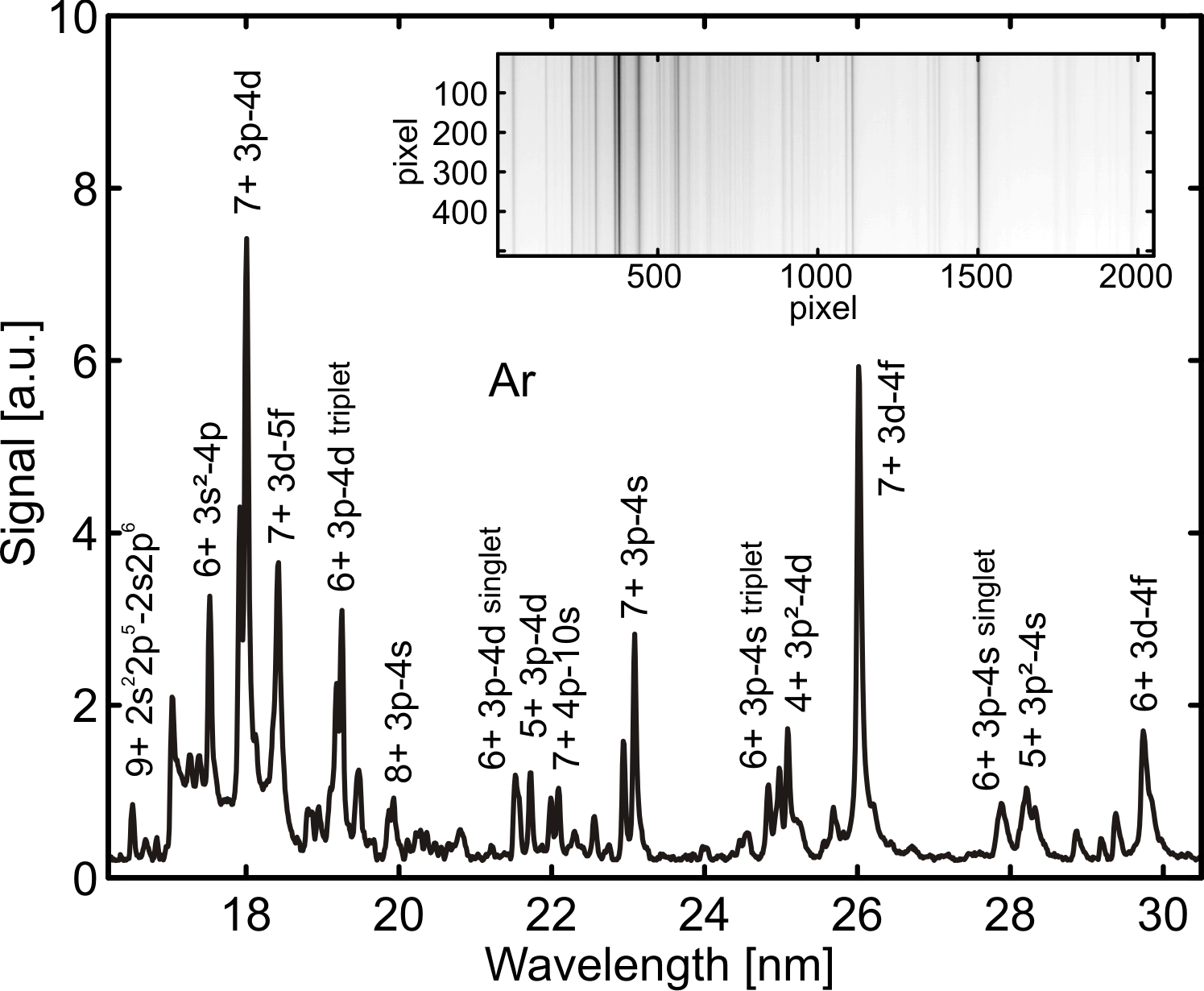}
\caption{EUV spectrum measured after exposure of argon microdroplets
  by intense ultrashort laser pulses at intensities of
  $8\cdot10^{14}$ W/cm$^2$. Several emission lines resulting from
  argon charge states Ar$^{4}$ to Ar$^{9}$ can clearly be
  resolved.\newline Inset: Raw data of the flat-field spectrometer.}
\label{fig:exp_Ar_spectra}       
\end{figure}

\begin{figure}[t]
\centering
\includegraphics[width=0.5\textwidth]{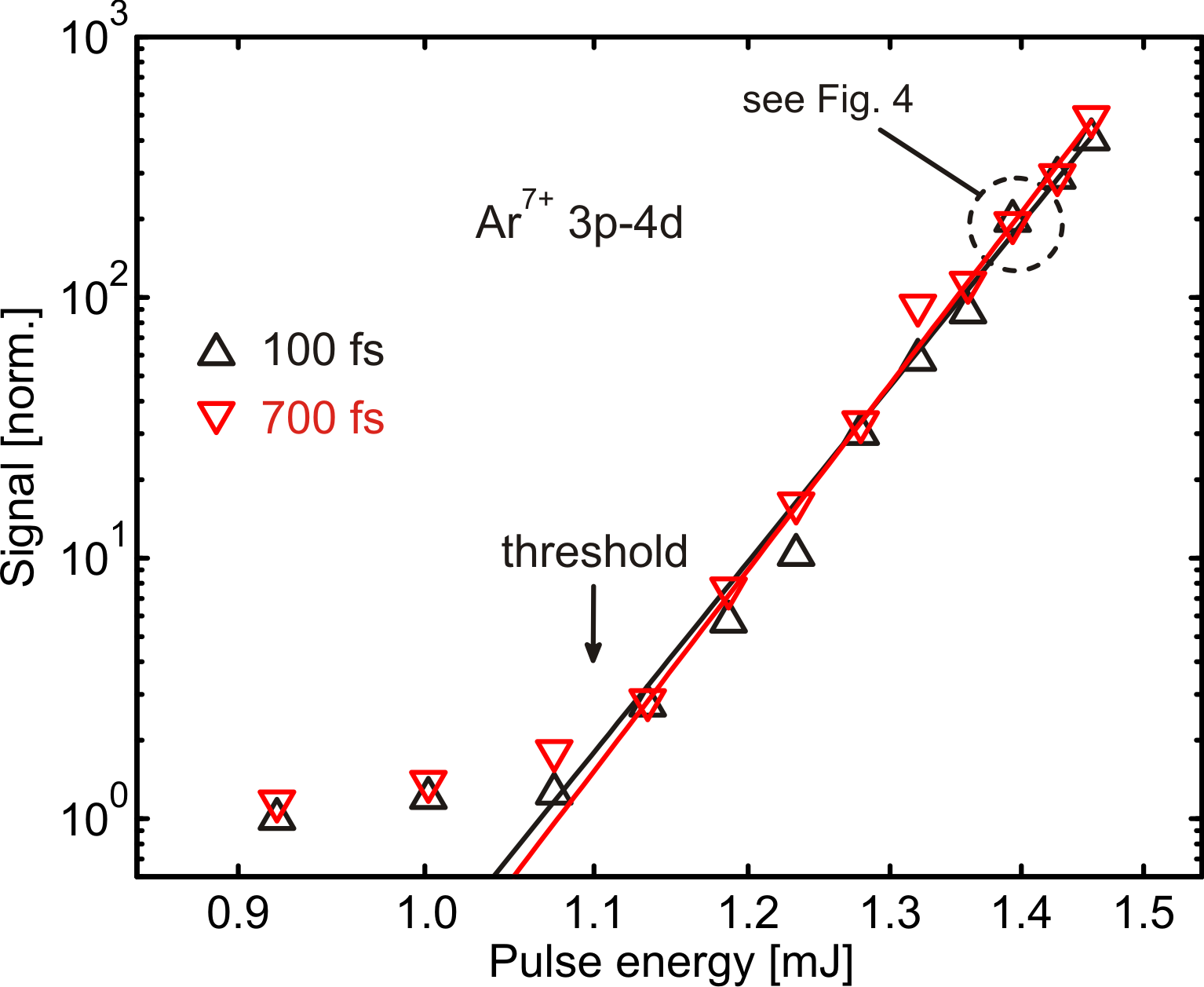}
\caption{(color online) Experimental Ar$^{7+}$ 3p-4d yields on a double logarithmic scale as
  function of the laser pulse energy for selected pulse durations. The data are normalized to the value obtained with 0.92\,mJ and 100\,fs. The pronounced nonlinear increase above laser energies of about 1.1\,mJ
  points at the highly nonlinear plasma dynamics. Similar findings have
  been obtained for all other transitions present in the spectrum in Fig \ref{fig:exp_Ar_spectra}.}
\label{fig:exp_Ar7_vs_Ep}       
\end{figure}

Fig. \ref{fig:exp_Ar_spectra} shows an EUV spectrum recorded after
exposure of microdroplets to 100\,fs laser pulses at intensities of
$8\cdot10^{14}$ W/cm$^2$. The characteristic line pattern can be
assigned to atomic transitions from Ar$^{4+}$ to
Ar$^{9+}$ \cite{ZweibackPRA1999,NIST}.  Note that under the given
intensity conditions, the laser field ionization of single atoms only leads
to charge states up to Ar$^{3+}$ \cite{AugstPRL1989} indicating that
anticipated processes like collisional ionization must
contribute. In the following we concentrate on the ion species Ar$^{7+}$ 
as it represents a characteristic and dominant fingerprint of the microplasma.
Fig. \ref{fig:exp_Ar7_vs_Ep} shows the normalized yields of the
Ar$^{7+}$ 3p-4d transition line depending on the pulse energy for pulse durations of 100\,fs and 700\,fs. A strong increase of the EUV
signal is observed beyond a 
threshold of 1.1\,mJ. 
At 700\,fs and 1.1\,mJ, the laser intensity matches the argon tunnel ionization threshold \cite{AugstPRL1989}, but is a factor of 7 higher for 100\,fs pulses. One obtains no significant change in the emission threshold behavior if shorter pulses are used. Charging of the droplet by only tunnel ionization is thus not sufficient to cause plasma EUV emission. 
 The strong nonlinear dependence around the EUV threshold points to an avalanche-like process whereby collisional ionization and inverse bremsstrahlung efficiently heat the target.
For both pulse durations, the yields enhance by more than two orders of magnitude, although the pulse energy only increases from 1.1 to 1.5\,mJ. 

\begin{figure}[t]
\centering
\includegraphics[width=0.5\textwidth]{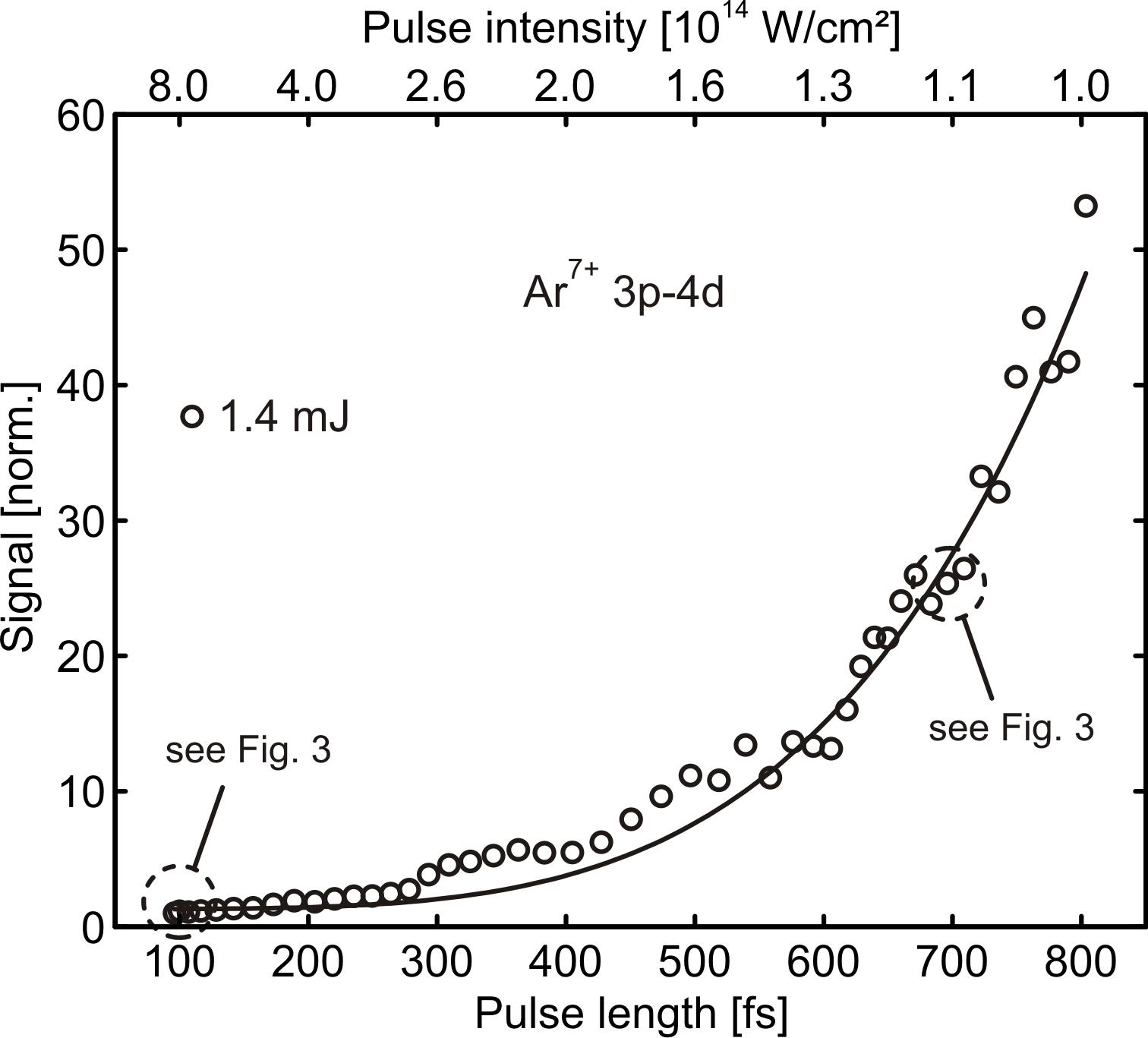}
\caption{Experimental dependence of the Ar$^{7+}$ 3p-4d EUV signal on
  pulse duration for a given pulse energy of 1.4\,mJ (see
  Fig. \ref{fig:exp_Ar7_vs_Ep}). The data are normalized to the pulse duration of 100\,fs. The yields increase with pulse length although the pulse intensity decreases (upper scale).}
\label{fig:exp_Ar7_vs_tp}       
\end{figure}

Whereas the plot in Fig. \ref{fig:exp_Ar7_vs_Ep} appears to suggest that the pulse duration  is of little importance, a closer look reveals a differentiated perspective.
Fig. \ref{fig:exp_Ar7_vs_tp} shows the dependence of the Ar$^{7+}$
3p-4d emission on the pulse duration for a given pulse energy of 1.4\,mJ. Apparently, an extended heating
period with stretched pulses at energies above the emission threshold is more effective even though the laser intensity decreases. 
We emphasize that similar findings have
been obtained in experiments on small clusters \cite{KoePRL99}.

\begin{figure}[t]
\centering
\includegraphics[width=0.5\textwidth]{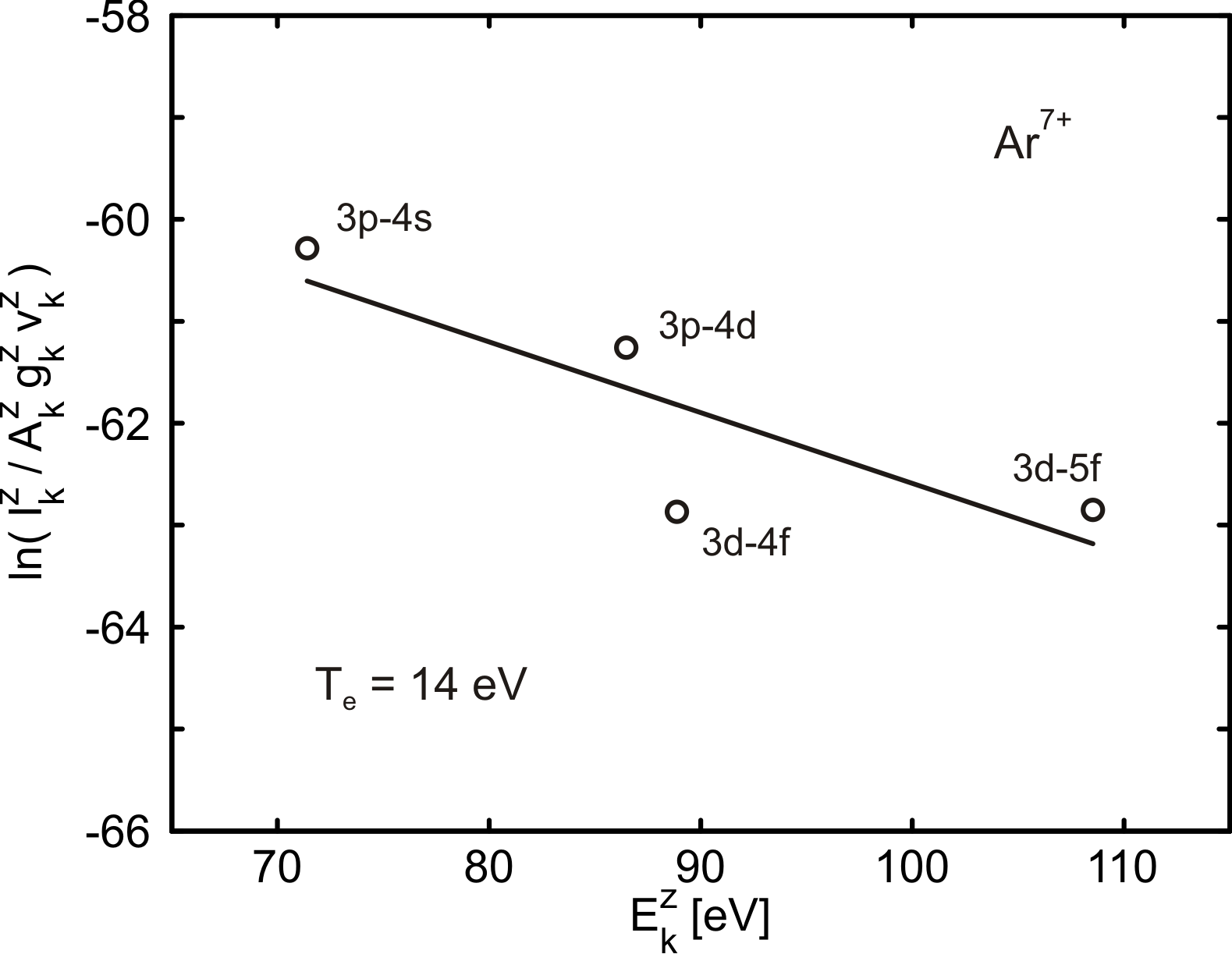}
\caption{Extraction of the electron temperature from a Boltzmann plot
  using the emission lines of Ar$^{7+}$. The corresponding spectrum
  has been recorded with a laser pulse energy of 1.4\,mJ and a pulse
  duration of 100\,fs.  From the experimental data $T_e$=14\,eV is
  deducted. Note that the transition Ar$^{7+}$ 4p-10s from Fig. \ref{fig:exp_Ar_spectra} is not included, as the Einstein coefficient is not available in this case. As well, determination from other emission lines, e.g. Ar$^{6+}$, could give a slightly shifted $T_e$. The Boltzmann plot method is sensitive only to the spatial area, where a certain ion species is abundant and contributes to the signal significantly.}
\label{fig:exp_bltzmnplt}       
\end{figure}

From the emission spectra obtained in the experiment, relevant
plasma parameters as for example the electron temperature $T_e$ can be
extracted. We apply the Boltzmann plot method to determine $T_e$ from line intensities \cite{Griem1997} assuming the microplasma to be at local thermodynamic equilibrium (LTE) at the time of the EUV emission. Briefly, the density $n_k$
of a certain ion species $k$ as function of $T_e$ can be
described by a Boltzmann distribution $n_k=n_0g_k / Z_k(T_e) \exp(-E_k
/k_BT_e)$, where $n_0$ is the total density, $g_k$ the
statistical weight, $Z_k(T_e)$ the partition function, $E_k$ the
upper energy level and $k_B$ the Boltzmann constant. The relative line
intensity $I_k^z$ of a certain transition $z$ in the charge state $k$ scales by $I_k^z\sim h\nu_k^z A_k^z n_k^z$,
with $\nu_k^z$ and $A_k^z$ the frequency and Einstein coefficients,
respectively.  One can derive a linear equation in the form of
$\ln(I_k^z/A_k^z g_k^z \nu_k^z) =-E_k^z/k_BT_e+const$. A linear regression taking
into account the emission lines of a given charge state allows to
extract $T_e$.  Fig.~\ref{fig:exp_bltzmnplt} shows the
Boltzmann plot analysis for Ar$^{7+}$ giving a value of $T_e$ of about
14\,eV ($1\pm20\%$) in this case. 
The accuracy in the determination of $T_e$ is comparable to other work in the warm dense matter regime, e.g. \cite{FaePRL10}.

\section{Computational results}

\begin{figure}[t]
\centering
\includegraphics[width=0.5\textwidth]{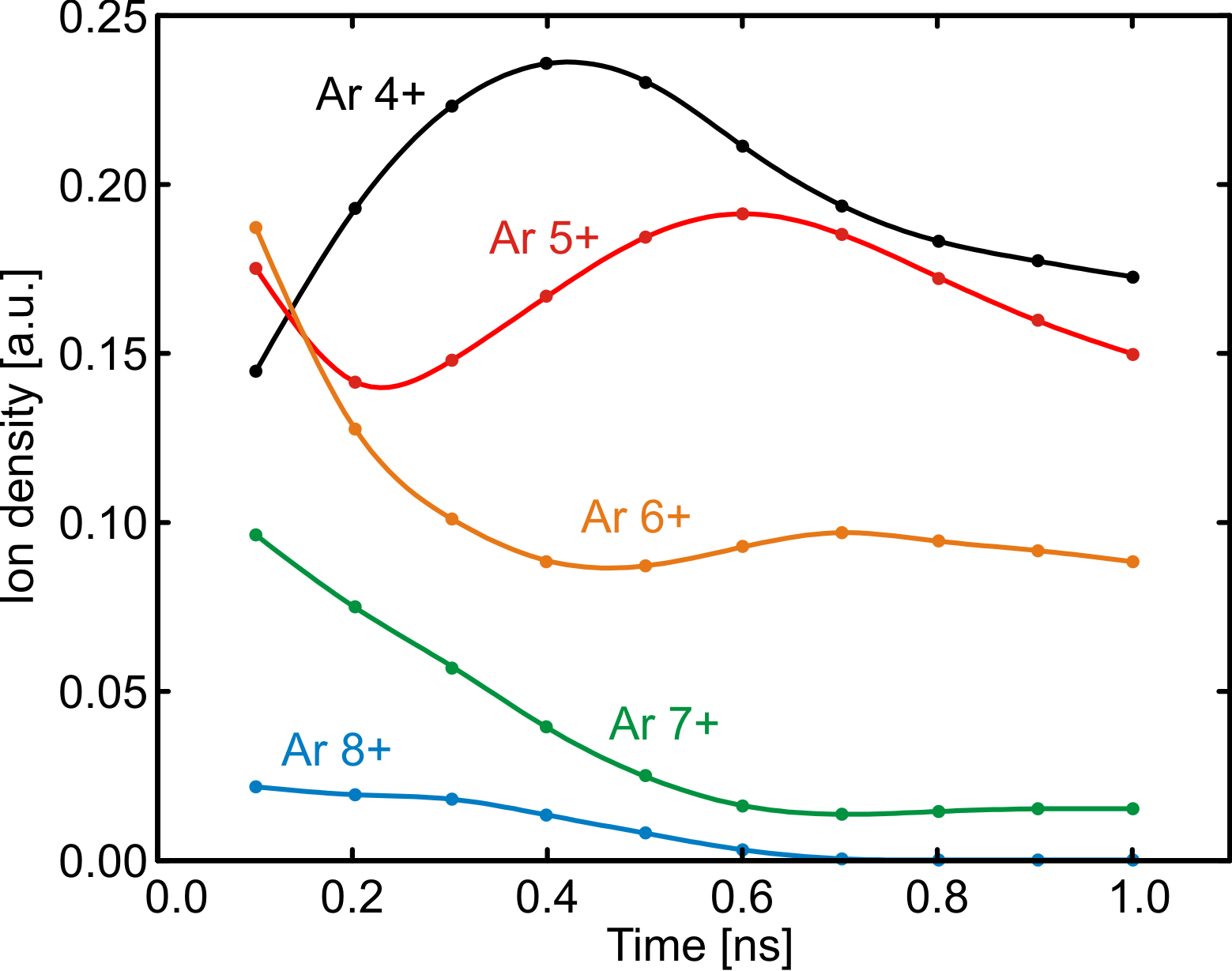}
\caption{(color online) Calculated temporal density distribution of argon ion
  species generated in the plasma layer for laser pulses with 100\,fs and 2.1\,mJ. (Time steps of 100\,ps are used. The curves are drawn to guide eyes.) Charge states from Ar$^{4+}$
  to Ar$^{8+}$ are shown. Ar$^{9+}$ is not observed in the
  simulation. Note that in the interaction with the laser pulse
  itself, up to Ar$^{3+}$ can significantly be produced by
  tunnel ionization.  The arise of higher charge states indicates the contribution of inverse 
	bremsstrahlung and collision ionization to the plasma heating. 
	The plasma dynamics lasts over nanoseconds after the laser pulse trigger at $t=0$.}
\label{fig:sim_nArs}       
\end{figure}

In the HELIOS calculations, argon droplets are considered with an initial diameter of
10\,$\mu$m and a mass density of $\rho_0=1.4$\,g/cm$^3$ irradiated by
intense laser pulses at 800\,nm center wavelength. Fig. \ref{fig:sim_nArs} shows the temporal ion density distributions $\rho_q$ of various charge states for laser pulses with 100\,fs pulse length and 2.1\,mJ pulse energy. The charge
state abundances are captured up to nanoseconds after the
laser pulse exposure at $t=0$. The ionization kinetics are caused by collisional excitation and 
recombination as well as collective processes in the plasma layer. In
the expansion, electron-ion recombination leads to a general decrease
of $\rho_q$. The short-term enhancement in the ion densities of
certain charge states after some hundred picoseconds can be traced back to
electron localization reducing the value of $q$ in higher charged
Ar$^{q+}$.  

\begin{figure}[t]
\centering
\includegraphics[width=0.5\textwidth]{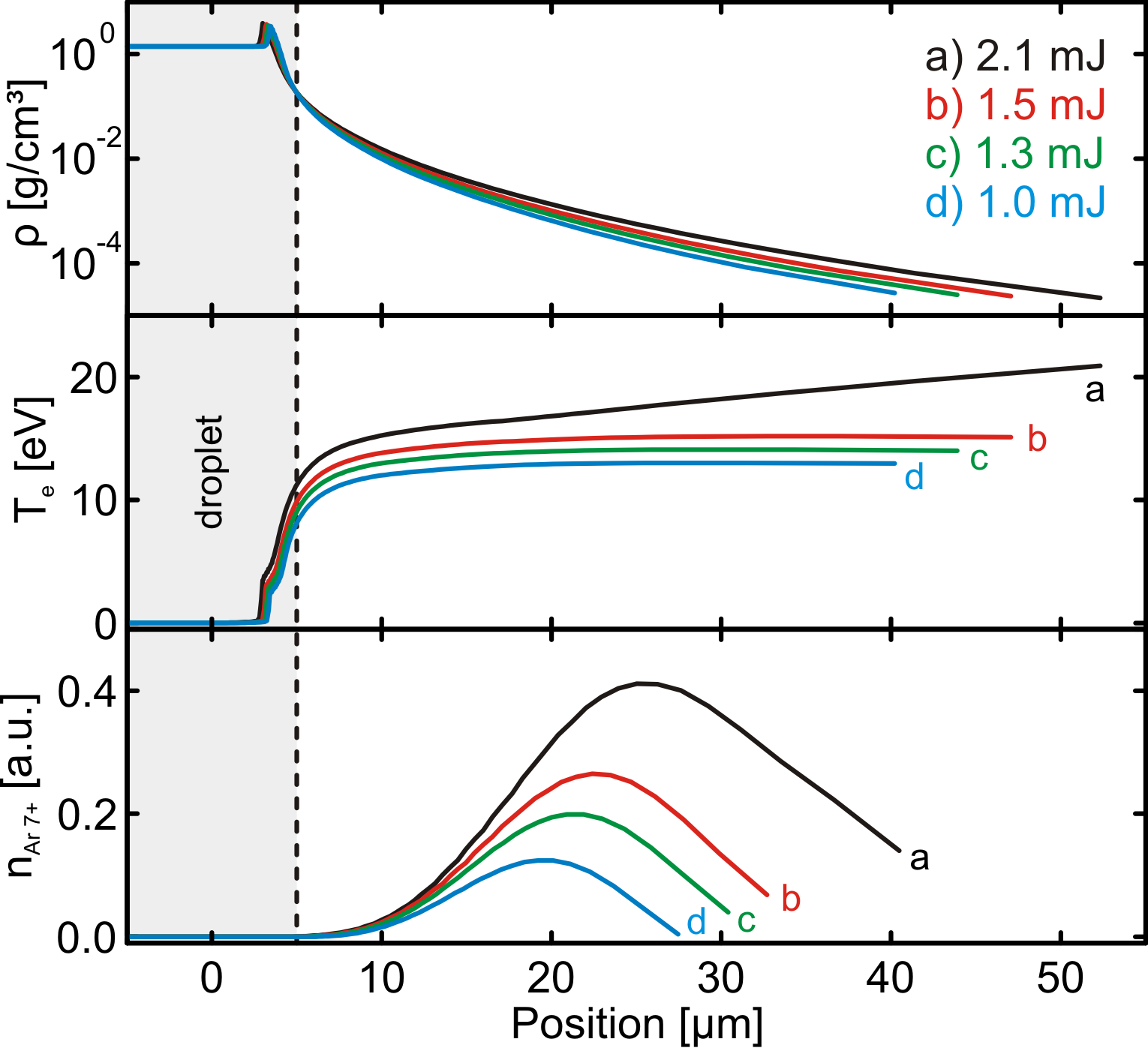}
\caption{(color online) Snapshots of spatial distributions of mass density $\rho$ (top), electron temperature $T_e$ (center) and Ar$^{7+}$ ionization fraction $n_{Ar^{7+}}$ (bottom) calculated for $t=100$\,ps after the laser pulse trigger. Different pulse energies at a constant pulse length of 100\,fs are used. The initially irradiated surface of the microdroplet is located at a position of 5\,$\mu$m, the plasma expands to the right.}
\label{fig:sim_rho_Te_nArs}       
\end{figure}

\begin{figure}[t]
\centering
\includegraphics[width=0.5\textwidth]{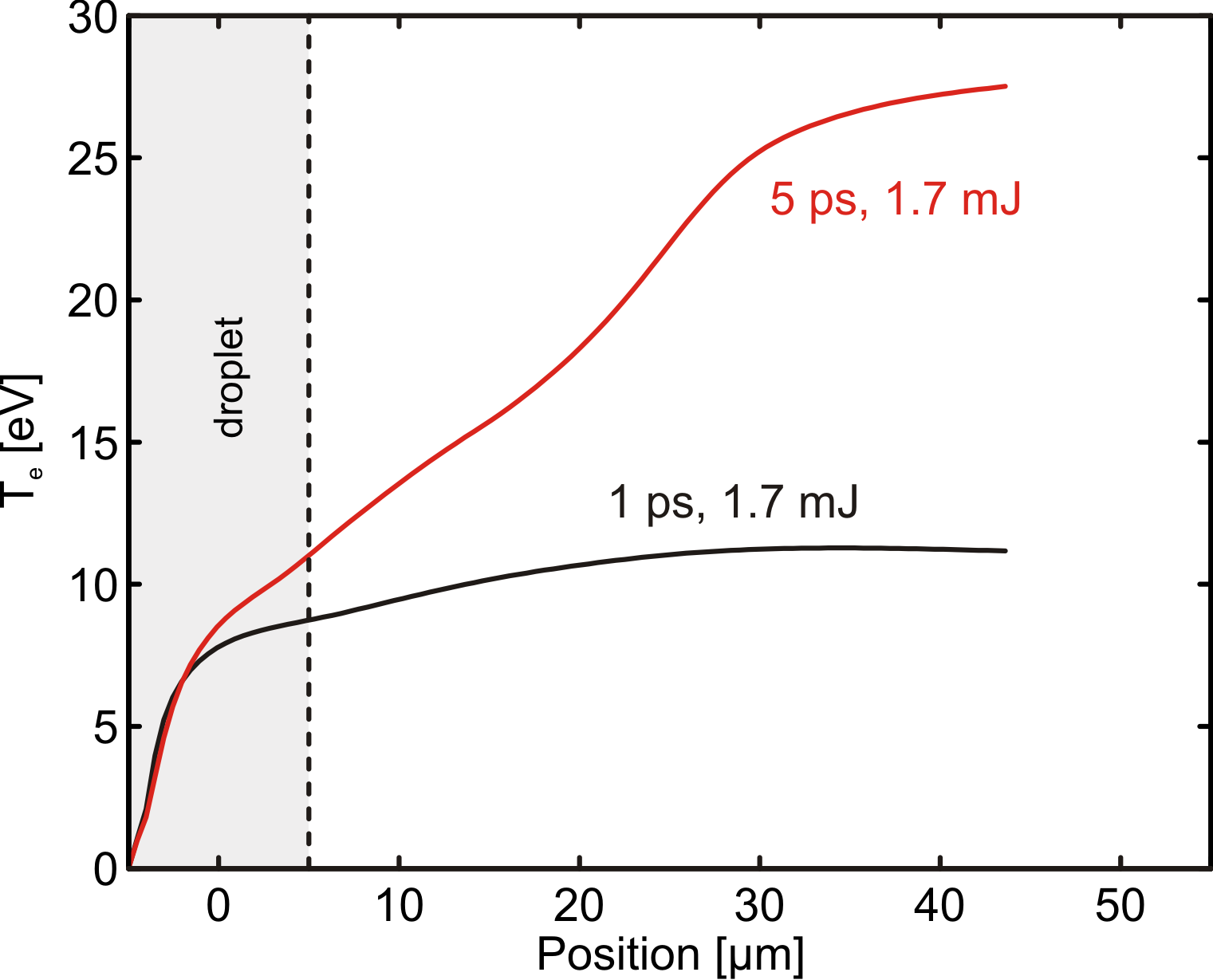}
\caption{(color online) Snapshots of calculated spatial distributions of the electron temperature $T_e$ at $t=1$\,ns for different pulse durations indicating that the interior of the droplet
  will be almost fully ionized. The influence of the pulse duration is greatest for the outer surface plasma layer.}
\label{fig:sim_Te_tp}       
\end{figure}

Fig. \ref{fig:sim_rho_Te_nArs} displays the spatial
evolution of the mass density and electron temperature as well as the ionization fraction of
Ar$^{7+}$ at $t=100$\,ps. Most of the inner part of the droplet (left
to dashed line) has a low temperature as the laser radiation is
shielded by the plasma layer.
As a consequence, the mass density inside the droplet remains at its initial value
$\rho_0$. A plasma compression is observed in the surface
region. A hot, highly ionized plasma layer extends over tens of micrometers outwards. 
The mass density drops - depending on the laser pulse energy - to values as low as $10^{-4}$\,g/cm$^3$ beyond 40\,$\mu$m.
The composition of the plasma varies with distance.
An enhanced abundance of Ar$^{7+}$ (Fig. \ref{fig:sim_rho_Te_nArs} bottom) is observed in a layer 
between 20 and 40\,$\mu$m approaching values of 10 to 40\,\% of the total density in the plasma plume. In general,
only a minor dependence of $T_e$ and $\rho$ on the pulse energy is
obtained for the considered range of parameters. The inner region of the microdroplet is still almost unaffected. However, in particular hot electrons penetrate and slowly heat-up the droplet core leading to an increasing ionization. 
After $t=1$\,ns, the droplet is almost fully ionized
as illustrated in Fig. \ref{fig:sim_Te_tp}. Interestingly and in
accordance with the experimental findings, see
Fig. \ref{fig:exp_Ar7_vs_tp}, longer pulses extend the heating period
and lead to an increase in the temperature of the microplasma and, thereby, enhanced population of higher ionization states. Whereas the inner region of the droplets is nearly unaffected by a variation of the pulse duration, the temperature near the surface and, especially, in the outer plasma region increases applying longer pulses.

\begin{figure}[t]
\centering
\includegraphics[width=0.5\textwidth]{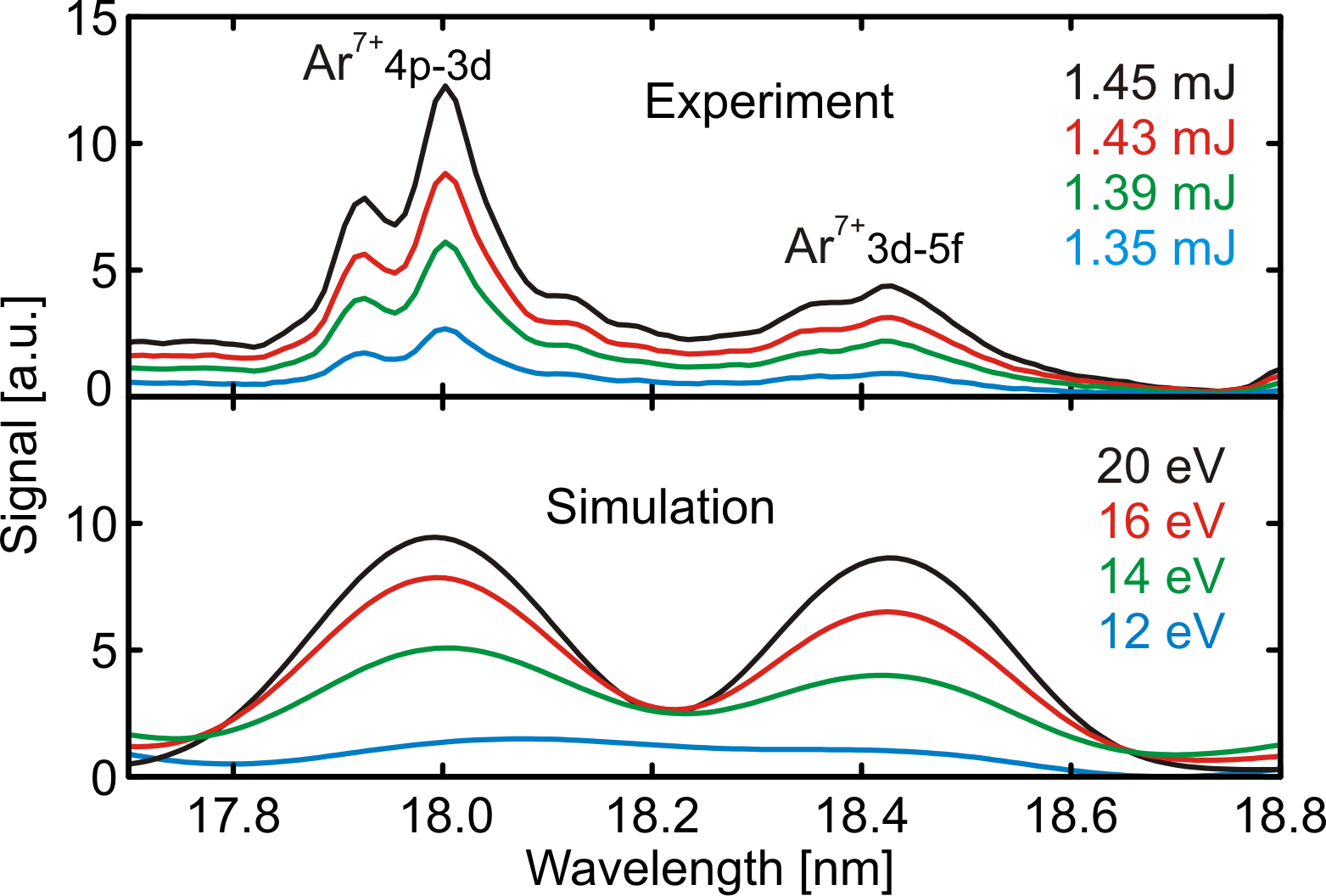}
\caption{(color online) Dependence of the signal of selected nearby emission lines
  (Ar$^{7+}$ 4p-3d and 3d-5f) on the plasma conditions. Experimental results for different laser energies (top) are compared with SPECT3D calculations for different electron temperatures (bottom). The experimentally observed enhanced line intensities with increased laser pulse energies can be traced back theoretically to an increase of temperature in the plasma layer.}
\label{fig:exp_sim_zoom_Ar7_spectra}       
\end{figure}

\begin{figure}[t]
\centering
\includegraphics[width=0.5\textwidth]{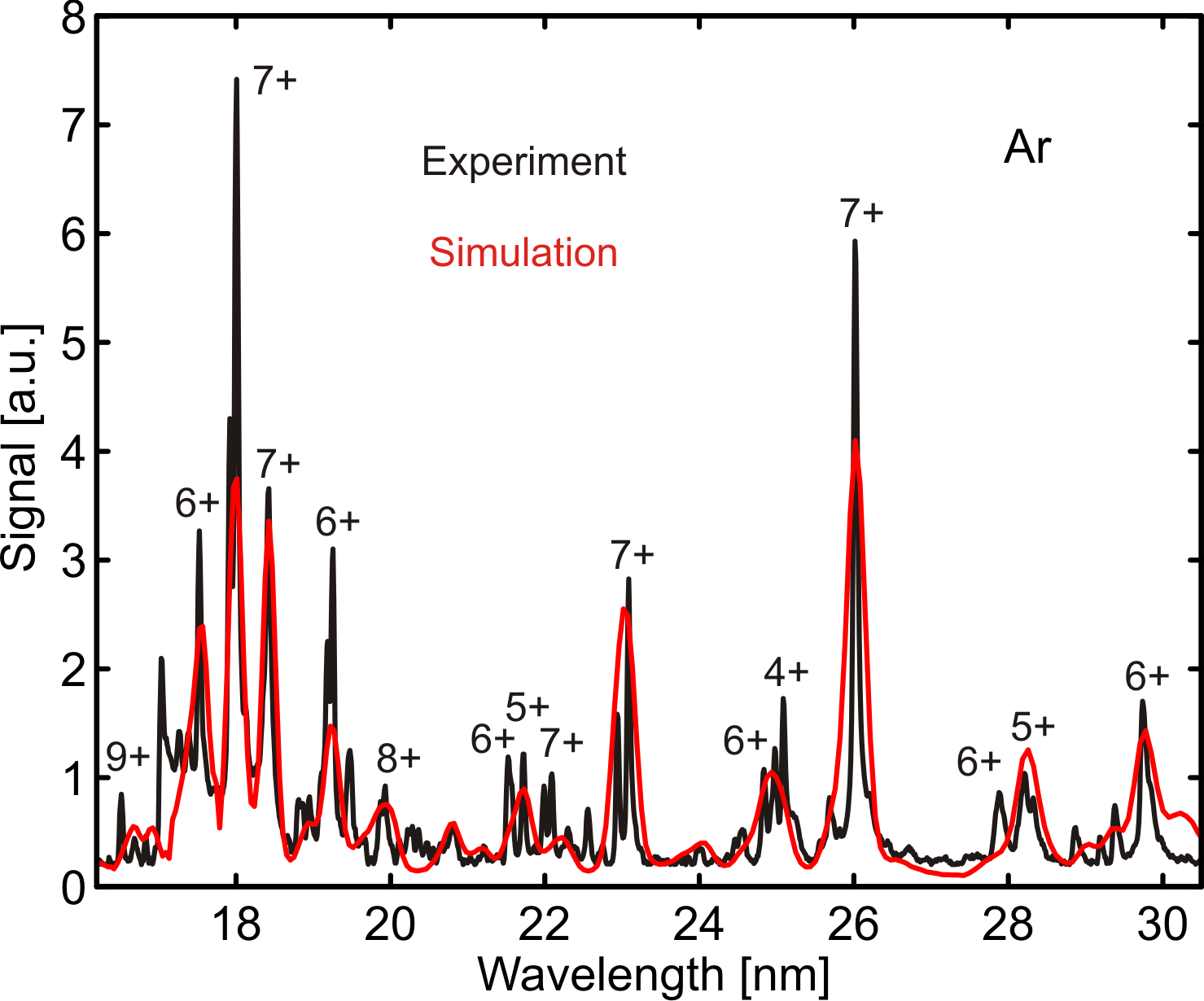}
\caption{(color online) Experimentally recorded EUV emission taken at 1.4\,mJ pulse energy (black) and calculated spectrum for an electron temperature of 16\,eV (red), respectively. Equal charge states and atomic transitions are observed and thus similar plasma conditions can be assumed.}
\label{fig:exp_sim_Ar_spectra}       
\end{figure}

Finally we like to compare the experimental findings with the results
obtained in the simulations. For this purpose the resulting soft x-ray
spectra using the SPECT3D software
package are computed~\cite{MacFarlaneHEDP2007}. For given $T_e$ the relative line
intensities are calculated and compared to the measurement, see
Fig. \ref{fig:exp_sim_zoom_Ar7_spectra}.  For a complete description, one should consider that the EUV
plasma emission will originate from the full period of the microplasma
expansion. Hence, only averaged values of, e.g., $T_e$ can be
extracted from the measurements.  Taking this into account, a good
agreement is found between experiment and theory as details of the
spectra match almost consistently, see Fig. \ref{fig:exp_sim_Ar_spectra}. The corresponding plasma electron
temperatures also agree although the calculations have been performed
using a one-dimensional hydrodynamic code which cannot reveal the full spatial evolution of the plasma.

\section{Conclusion}

Soft x-ray radiation from argon microdroplets irradiated by intense
femtosecond pulses has been analyzed near the EUV emission threshold
in a joint experimental and theoretical study. At given pulse energy,
charging of the droplets can be controlled by the pulse duration. The longer impact time supports plasma
heating leading to an enhancement of the EUV emission. Hydrodynamic simulations reveal that plasma formation initially
takes place on the surface of the droplet. In the expansion of the plasma
layer spatial and temporal charge state distributions change up to
nanoseconds. Whereas the outer plasma cools down in the adiabatic
expansion, the inner region of the droplet heats up on a comparable
time scale. 
The experimental spectrum contains contributions from the full development of the microplasma expansion. 
Nevertheless average electron temperatures extracted
from the EUV emission signal are found to be in good agreement with 
hydrodynamical (HELIOS) and collisional-radiative (SPECT3D) simulations. In further work the analysis of the plasma parameters may be improved by applying the Saha-Boltzmann distribution instead of the Boltzmann distribution.

\section{Acknowledgment}

We gratefully acknowledge financial support by the DFG within the SFB 652 and by the BMBF
within the FSP 302.

\clearpage 

\singlespacing 
\bibliography{lib}
\bibliographystyle{iopart-num}

\end{document}